\documentclass{aa}

\usepackage{epsfig,amssymb}

\begin{document}
%
\title{Quiet \object{Sun} coronal heating: statistical model}

\subtitle{}

\author{V.~Krasnoselskikh
           \inst{1}
           \and
        O.~Podladchikova
           \inst{1}
           \and
        B.~Lefebvre\inst{2}
           \and
        N.~Vilmer
           \inst{3}
       }

\offprints {V.~Krasnoselskikh,\\
email: vkrasnos@cnrs-orleans.fr}

\institute{ LPCE/CNRS, 3A Av. de la Recherche Scientifique,
            45071 Orl\'eans, France
\and
            ESST, Kyushu University, 6-1 Kasuga-Koen,
            Fukuoka 816-0943, Japan
\and
            DASOP, Paris-Meudon Observatory,
            5~pl. J.~Janssen, 92195 Meudon, France
          }

\date{Received 1 February 2000 / accepted  .........}

\authorrunning{V.~Krasnoselskikh et al.}
\titlerunning{Quiet Sun coronal heating}

\abstract{Recent observations of Krucker \& Benz
(\cite{Krucker98}) give strong support to Parker's hypothesis
(\cite{par}) that small scale dissipative events make the main
contribution to quiet Sun coronal heating. They also showed that
these small scale events are associated not only with the magnetic
network, but also with the cell interiors (Benz \& Krucker,
\cite{Benz98}). Taking into account in addition the results of the
analysis performed by Priest with co-authors (\cite{pr1}) who
demonstrated that the heating is quasi-homogeneous along the arcs
we come to the conclusion that the sources driving these
dissipative events are also small scale sources. Typically they
are of the order of or smaller than the linear scale of the events
observed, that is $<$ 2000 km. To describe statistical properties
of quiet Sun corona heating by microflares, nanoflares, and even
smaller events, we consider a cellular automata model subject to
uniform small scale driving and dissipation. The model consists of
two elements, the magnetic field source supposed to be associated
with the small scale hydrodynamic turbulence convected from the
photosphere and local dissipation of small scale currents. The
dissipation is assumed to be provided by either anomalous
resistivity, when the current density exceeds a certain threshold
value, or by the magnetic reconnection. The main problem
considered is how the statistical characteristics of dissipated
energy flow depend upon characteristics of the magnetic field
source and on physical mechanism responsible for the magnetic
field dissipation. As the threshold value of current is increased,
we observe the transition from Gaussian statistics to power-law
type. In addition, we find that the dissipation provided by
reconnection results in stronger deviations from Gaussian
distribution.
}

\maketitle

\section{Introduction}  

Although photospheric convection may supply enough energy to heat
the coronal plasma, the way how energy is supplied and dissipated
in the corona is not completely understood (e.g. Priest et al.,
\cite{pr1}; Einaudi \& Velli, \cite{Einn}). Parker (\cite{par})
suggested that the solar corona could be heated by the dissipation
at many small scale tangential discontinuities arising
spontaneously in the coronal magnetic fields braided and twisted
by random photospheric footpoint motions. To emphasize these
events, he introduced the notion of nanoflares. This idea
stimulated the intensive search of observational signatures of
microflares and nanoflares as well as many theoretical
developments on the contribution of small scales to energy
dissipation in the solar corona. The solar flares, most energetic
impulsive phenomena in active regions were observed and studied in
detail for quite a long time. On a large scale, it is well known
that the most energetic dissipative events occur in the vicinity
of tangential discontinuities at the borders of the photospheric
dipolar regions like bright spots. These sources create strong
currents resulting in reconnection and flare-like events. Such
eruptive events are related to magnetic energy releases, sudden
changes of the magnetic field topology, and plasma heating.
Therefore Parker's idea about solar corona heating by nanoflares
used quite similar physical concepts on the nature of dissipative
phenomena, but it pointed out the importance of small scales.

The smaller scale phenomena, microflares, were first detected in
soft X-rays in a balloon experiment by Lin et al. (\cite{Lin}).
The development of new instrumentation allowed to perform
multi-wave satellite (YOKOH, SOHO, TRACE) and ground based (VLA)
high-resolution observations of smaller scale (about thousand
kilometers) lower energy phenomena. They were observed in active
regions but also in the quiet regions of the Sun and in coronal
holes. The small intensity flares observed in active regions were
quite rare and not powerful enough for heating (Shimizu et al.,
\cite{Shimizu94}). These observations were performed in the
regions of high level of background flux and strong stray light.
The very same microflares can be more easily detected in the quiet
corona. Soft X-ray observations (Benz et al., \cite{Benz97}) and
EUV observations (Harrison, \cite{Harrison97}) have revealed
enhanced emission and thus intense heating above the magnetic
network. Similar phenomenon that form small X-ray jets at the limb
was reported by Koutchmy et al. (\cite{Koutchmy97}). It was found
that the number of observed events increases with the sensitivity.

Another way to shed some light on the important problem concerning
the characteristic scales of the major dissipative events consists
in statistical study of different observed parameters. The major
objective of such a study is to obtain the Probability Density
Function of energy in a wide range of energies.

For regular flares that occur mainly in active regions Crosby et
al. (\cite{Crosby93}) have found that the total energy in the
flare electrons observed in hard X-ray bremsstrahlung has a power
law distribution $f(E) \sim E^{ \alpha}$ with index $\alpha =
-1.53\pm 0.02$. But the energy supplied by the flares in the
active regions is not sufficient for the corona heating.

The crucial factor that determines the characteristics of the
heating process and its scales is related to the power law index
of the PDF. If it is larger than minus two the major heating takes
place in the intensive large scale dissipative events. If it less
than minus two it occurs in small scales. The reason is that to
provide the explanation of the efficient coronal heating in small
scales it is necessary to have an excess in the occurrence rate of
small scale events. An important result that supports the
hypothesis of Parker was reported by Krucker \& Benz
(\cite{Krucker98}) who have found from the Yohkoh / SXT
observations that the PDF dependence upon the energy has a power
law distribution in the energy range $10^{24}$--$10^{26}$ ergs
with the index about $-2.59$. This result was obtained assuming
that the flaring region has a constant height. The authors not
only concluded that the weak flaring events rule the heating
process, but they also noticed that the brightest small scale
microflares lie above enhanced elements of the magnetic network,
and the denser ones exhibit a higher level of fluctuations. This
means that the corona is continuously replenished by underlying
chromospheric material that has been heated to coronal
temperatures. Mitra \& Benz (\cite{Mitra01}) have discussed the
same observations but supposing that the height varies
proportionally to the square root of the area and have shown that
the index becomes a little larger but still is smaller than minus
two. This result was confirmed by Parnell \& Jupp
(\cite{Parnell00}), who estimated the index to be between $-2$ and
$-2.1$ making use of the data of TRACE. However, Aschwanden et al.
(\cite{Asch00}) using also the data of TRACE have found
significantly different value of this index $-1.80$ suggesting
that there may not be enough energy in microflares or nanoflares
to heat the entire quiet corona. These last authors casted doubt
on the hypothesis that the heating takes place in small scales,
which according to them remains controversial.

It is also worth mentionning that Benz \& Krucker (\cite{Benz98})
and Berghmans et al. (\cite{Berg98}) have noticed that the heating
events occur not only on the magnetic network boundaries but also
in the cell interiors. They came to the conclusion that these
events have a quite similar nature.

The idea of small scale intermittent behavior of dissipation was
elaborated in theoretical models by several authors. Using an MHD
approach, Einaudi \& Velli (\cite{einaudi82}) investigated the
dependence of the dissipated energy upon the scales. Different
kinds of ``fragmentation of dissipation'' were discussed by Vlahos
(\cite{vlah}). Einaudi et al. (\cite{einaudi96}) and Georgoulis et
al. (\cite{manolis}) simulated 2D MHD systems driven by large
scale forces and with diffusive dissipation. Their 2D simulations,
driven by a pair of large scale vortices with random phases and
amplitudes, showed that the spatial average of the dissipated
power had a non-Gaussian statistics. After subtraction of the
Gaussian component, the dissipated events were shown to have a
scaling law distribution.

The observed power-law distributions of the energy flux for the
largest dissipative events, flares, and microflares inspired the
development of models based on the idea of Self-Organized
Criticality (SOC) (Lu \& Hamilton, \cite{Lu-H}; Lu et al.,
\cite{Lu}; Vlahos et al., \cite{vlah2}; Georgoulis et al.,
\cite{Georg3}; Georgoulis \& Vlahos, \cite{Georg1,manolis}). Lu et
al. ( \cite{Lu}) and Georgoulis et al. (\cite{Georg4}) compared
the predictions of the models with the observations of the flares.
SOC has appeared as a paradigm for slowly driven complex systems,
which exhibit power-law-type relaxation events and correlations of
infinite range (Jensen, \cite{Jensen}). These models are similar
to the original "sandpile" model (Bak et al., \cite{Bak}), but the
sand is replaced by the magnetic field flux. The dissipation
occurs through small scale reconnection, when a "current," which
is defined in terms of the difference between the field in a
particular cell and the average over all nearest neighbors,
exceeds some threshold. In these models, a perturbation in a
single cell can trigger quite a large cascade of reconnections.
This results in a power-law energy distribution of dissipative
events and power-law spatial correlations. However, one should
take care interpreting these results because a small grid size can
result in an artificial form of the PDF's obtained. A heuristic
justification of such models based on 2D MHD equations was
proposed by Vassiliadis et al. (\cite {vas}). It is also worth
mentioning that, using a shell model of MHD turbulence, Boffetta
et al. (\cite{Bof}) have demonstrated that there can exist
alternative reasons for a scaling law to appear.

Our model is also based on the idea of cellular automata. Its
difference with the previous ones consists in the assumption that
not only the dissipation takes place in small scales but  the
magnetic field sources are also of small scale. This is in
agreement with the idea of Benz and Krucker (\cite{Benz98}) that
the heating can take place on the level of chromosphere, implying
that the magnetic field structures have the scales of the order or
less than its height. This feature results from the analysis of
Priest with co-authors (\cite{Priest98}) who have shown that the
heating is quasi-homogeneous over the height of the magnetic loop.
Other indication about the possible role of the small-scale
sources follows from observations of Abramenko et al.
(\cite{abramen}) of the evolution of the vertical component of the
current helicity of an active region magnetic field during a
transition from a low flaring state to an enhanced one. They found
that the reorganization of the vortex structure of the
photospheric magnetic field occurs, when small scale vortices
re-organize into larger scale vortices, suggesting that an inverse
helicity cascade develops.

Although the footpoints of the loops are known to move randomly,
the statistical properties of their motions are not known in
detail. These sources create strong currents resulting in
reconnection and flare-like events. Such eruptive events are
related to magnetic energy releases, sudden changes of the
magnetic field topology, acceleration of particles and plasma
heating. Another possible source of energy dissipation and coronal
heating is provided by an anomalous resistivity resulting from
current driven instabilities which are developed when the currents
exceed a certain threshold value. It provides Joule-like
dissipation in collisionless plasma. This results in a relatively
slow heating, but does not give rise to particle acceleration as
for the previous dissipation mechanism. Moreover it may occur
everywhere in the turbulent shear flows and not necessarily on the
boundaries of the magnetic field network.

Another important feature that makes the difference from previous
studies of CA type models is the homogeneous distribution of the
magnetic field sources. The problem to be addressed in the present
work is whether the small scale magnetic fields generated in (or
convected from) the photosphere or chromosphere (mainly at the
boundaries between granulas, but also inside them) can make a
significant contribution to the coronal heating.

We are mostly interested in the scales typical for nanoflares and
even in smaller ones. In this case, the small scale sources of the
magnetic field and the energy dissipation can have comparable
characteristic spatial scales. Thus the inertial interval as well
as the energy cascade are not so important in such a model as for
the conventional Alfv\'{e}nic or MHD turbulence. But the system
considered can possess the property of ``inverse cascade'',
thereby influencing the structure of the larger scale magnetic
fields as is supposed to be the case in the SOC-type systems. In
this paper we concentrate on the study of the dissipated energy
flow, in particular, on the dependence of statistical
characteristics of dissipated energy flow upon the characteristics
of the magnetic field source, and on the physical mechanism of the
magnetic field dissipation.

\section{Small scale driving and dissipation}   

\begin{figure}[!htb]  
\centerline{\epsfig{file=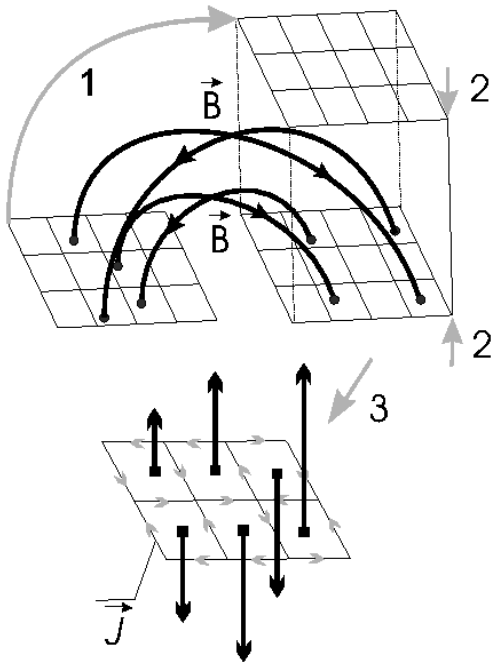,width=6.5 cm}}
\caption{A schematic representation of the procedure used in our
         model to simplify magnetic field configuration.
        }
\label{ARK}
\end{figure}

We consider hereafter the sand-pile-type cellular automata, where
sand piles are replaced by a magnetic field flux. In the model,
the relaxation process consists in the dissipation of the magnetic
field energy by means of the dissipation of currents, when their
magnitudes exceed a certain threshold value, or due to the
reconnection process.

In MHD approximation the evolution of the magnetic field is governed by
\begin{equation}
\frac{\partial\vec{B}}{\partial t}
=\nabla\times(\vec{u}\times\vec{B}) +
\mathrm{dissipative\ term},
\label{B}
\end{equation}
where $\vec{B}$ is the magnetic field, $\vec{u}$ is the fluid
velocity of the plasma. Without the dissipative term, this
equation describes the motion of the magnetic field lines frozen
into the plasma. In this case, the reconnection is forbidden. When
a finite conductivity $\sigma$ is taken into account in the form
of Ohm's law
$\vec{j}=\sigma\left(\vec{E}+\vec{u}\times\vec{B}\right)$, where
$\vec{j}$ is the current density, it gives rise to the diffusion
effect. In Eq.~(\ref{B}) the effect is described by a dissipative
term $\eta\Delta\vec{B}$, where $\eta=1/\sigma\mu_0$ and $\mu_0$
is the magnetic permeability of the free space. The diffusion time
$t_\eta=l^2/\eta$ is generally large for all scales but small
where it is comparable with the typical time for fast
reconnection. In this case it has the same order of magnitude as
an Alfv\'en time $t_A=l/V_A$, where $V_A$ is the Alfv\'en
velocity. It is worth mentioning that at small scales other
dissipative effects can become important (see, e.g., Biskamp
\cite{Bisk}).

In the present study, we use a simplified cellular automata model
mimicking Eq.~(\ref{B}).

Assume that the curvature of the magnetic field lines is
negligible. Then, in each cubic column only the upper and lower
faces should be considered, i.e., magnetic field lines are
straight and perpendicular to the surface. In this case the
equation $\nabla\cdot\vec{B}=0$ is automatically satisfied (see
Fig.~\ref{ARK}). As a result, the fields on the lower and upper
faces are equal, thus only one face will be considered. The 2D
surface is then split into cells. In most of the simulations
presented in the paper, grids of $200\times200$ cells are used
(unless otherwise stated). The boundary conditions are chosen to
be periodic.

As we have mentioned already in the introduction Krucker \& Benz
(\cite{Krucker98}) have found that the main heating occurs in the
small scale bright points. Making use of the multi-wavelength
analysis Benz \& Krucker (\cite{Benz99}) have shown that the
temporal sequence of observations of different wave emissions is
similar to the one in the large scale magnetic loops. They came to
the conclusion that the physical mechanisms of the energy release
are similar in these two cases. It follows then that the
dissipation process in the second case is also associated with the
magnetic loops, but of a small scale. Moreover, they have noticed
that the heating events occur not only on the boundaries of the
magnetic network but in the interiors of the cells also. The
comparative analysis of the model predictions for the plasma
heating in the magnetic loop due to the distributed energy source
with observations performed by Priest with co-authors
(\cite{Priest98}) led to the conclusion that the heating is
quasi-homogeneous along the magnetic loop. This means that the
heating process does not occur in the close vicinity of the foot
points but rather in the whole arc volume. If one will put
together these facts, it follows that the characteristic spatial
scale of the magnetic field loops which supply the magnetic field
dissipated is of the same order as the characteristic scale of the
dissipation. Thus we may conclude that not only the dissipative
process, but also the sources have small characteristic length.
Another conclusion from observations is that the sources are
distributed quite homogeneously in space. This reasoning leads us
to the choice of the sources and dissipation mechanisms used in
our model that we describe further.

Three kinds of sources with slightly different statistics are
considered.

\subsection{Source terms}   
\begin{figure}[!htb]  
\centerline{\epsfig{file=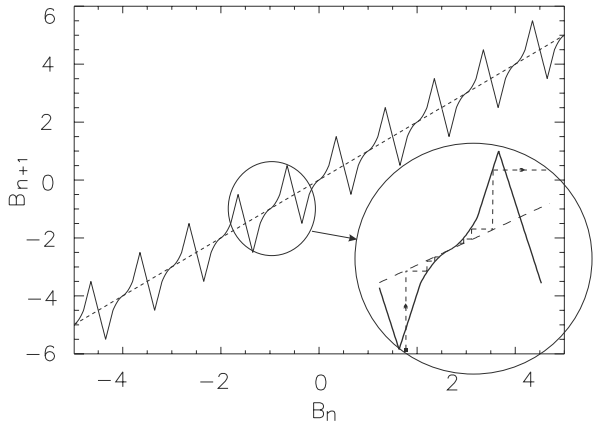,width=7.8cm}}
\caption{Graphical representation of the Geisel map (solid line).
         The fixed points of the map correspond to the intersections
         of the graph with the straight line $B_{n+1}=B_n$ (dashed
         line).
        }
\label{Geiselmap}
\end{figure}

We investigate the statistical behavior of the system driven by a
random and turbulent unipolar/dipolar sources. These sources are
used to model the effects of turbulent magnetic field convection
described by the first term on the right-hand-side in
Eq.~(\ref{B}).
\begin{itemize}
\item{\bf An unipolar random source.}
The simplest source of magnetic energy is an uncorrelated process
of zero mean, the values $\langle\delta B\rangle=0$ of which are
randomly chosen from the set $\{-1,0,1\}$, all the values being
equiprobable. In each time step the action of the source consists
in adding random numbers from the set mentioned above to the
previous values in the cell. The numbers are independently chosen
for each cell. This procedure automatically ensures that
$\langle\delta B\rangle=0$ for each cell.

\item{\bf A dipolar random source.}
Such a source can be made dipolar by dividing the grid into two
parts. For the positive and negative parts of the grid
the random numbers $\langle\delta B\rangle=0$ are chosen from the sets
$\{-0.5,0.5,1.5\}$ and $\{-1.5,-0.5,0.5\}$, respectively.

\item{\bf A chaotic source.}
Turbulence is certainly not a completely random process, and some
of its aspects can be simulated using deterministic models. The
Ulam map provides one of the simplest examples of a generic
chaotic system with quadratic non-linearity (see, e.g., Frisch,
\cite{Frisch}). The source $\delta B$ in each cell evolves
according to $$ \delta B_{n+1}=1-2(\delta B_n)^2, $$ where the
initial values of $\delta B_0$ are randomly chosen from the
interval $[-1,1]$. In this case, all other $\delta B_n$ belong to
the same interval. The action of the source is similar to the
above described for random sources.

\item{\bf A Geisel map source.}
In physical systems, the source term may depend on the value of
$B$ itself. When the dissipation is absent, the magnetic field in
each cell evolves according to the map $$ B_{n+1}=f(B_n). $$ The
initial values of $B_0$ are randomly chosen from the interval
$[-0.5,0.5]$. The systems like that are usually called Coupled Map
Lattices (CML) (Kaneko, \cite{kan}) rather than cellular automata.
We use the source based on the Geisel map (Geisel \& Thomae,
\cite{Geisel}) shown in Fig.~\ref{Geiselmap}. The fixed points of
this map that are defined by $B_n=f(B_n)$ are metastable. As a
result, an intermittency develops in the system, i.e., the
dynamics is slow and regular in the vicinity of the fixed points
of the map and fast and chaotic otherwise. As a consequence, the
map that describes the time evolution of the magnetic field in
each particular cell exhibits the behavior similar to anomalous
(non-Brownian) diffusion ,i.e. $$ \langle B^2\rangle\propto
t^\alpha,\quad\alpha<1. $$ The hypothesis has been suggested that
when the turbulence consists of convective cells, magnetic field
lines time evolution in each cell is similar to a subdiffusive
behavior ($\alpha<1$), which is, however, more complex than
described above.
\end{itemize}

\subsection{Dissipation criteria}   

The magnetic field dissipation provides the conversion of the
magnetic energy into thermal energy and ensures the coupling
between the magnetic field elements in our model.
Phenomenologically, reconnection can be treated as a dissipation
of small scale current sheets when the current density exceeds a
certain threshold value (Somov \& Syrovatsky, \cite{Som};
Syrovatsky, \cite{Syrov,Syrov2}).

If we neglect the displacement current, the current density can be
calculated from Maxwell-Amp\`ere's law
$\nabla\times\vec{B}=\mu_0\vec{j}$,
the finite-difference form of which can be written as
\begin{eqnarray*}
\left(
\begin{array}{c}
  j_x \\ j_y
\end{array} \right) = \frac{1}{\delta\mu_0} \left(
\begin{array}{c}
B\left(x,y\right)-B\left( x,y+\delta \right) \\
B\left(x+\delta,y\right)-B\left( x,y\right)
\end{array}
\right),
\end{eqnarray*}
where $\delta$ is the grid increment. For simplicity we let
$\delta=1$ and $\mu_0=1$. Currents are thus computed as local
gradients, and supposed to be carried along the borders between
the cells. It is seen that the discrete analog of the current
continuity equation $\nabla\cdot\vec{j}=0$ holds, i.e., the sum of
incoming and outgoing currents is equal to zero at each node of
the grid.

The mechanisms for the current dissipation are of two types.
\begin{itemize}
\item {\bf Anomalous resistivity},
arises as the result of an instability when the electric current
exceeds some critical value (which may be sometimes very small),
and results into an increase of the plasma's resistivity. This
phenomenon describes the phenomenon of pulse and energy exchange
between electrons and ions, or between different groups of
particles of the same nature by means of plasma turbulence in
collisionless plasma (Galeev \& Sagdeev, \cite{galeev,galeev1}).

When the electric current in plasma exceeds certain threshold
value plasma instabilities can be excited. As a result of the
instability development the waves are generated (e.g. drift waves,
or ion-acoustic, or low-hybrid waves). The development of the
instability gives rise to the absorption by generated waves of a
part of the electron's energy and pulse which is partly
redistributed to ions. This results to a dissipation of the
current quite similarly to Joule heating (see Appendix A for more
details). Finally, it is worth noting that anomalous resistivity
does not require any particular configuration of the magnetic
field, besides the strong current, and may occur even in the
presence of parallel magnetic fields pointing in the same
direction, as for example in coronal holes or cells interiors. In
our model we assume that the currents are completely annihilated
whenever they exceed a certain threshold, $$ |j|\geq j_{\max}. $$
\item {\bf Reconnection} is generally understood
as a relatively sudden change from one equilibrium state to
another, implying a change in magnetic field's topology,
accompanied by a transformation of magnetic energy into energy of
particules (Priest \& Forbes, \cite{pf:00}). In its ''primary
form'' the stationary reconnection process represents the
dissipation of the magnetic field in the vicinity of the so called
X-points of the magnetic field configuration or in the vicinity of
a current sheet. That can be interpreted as the rising and
``burnout'' of thin small scale current sheets separating domains
with oppositely directed magnetic fields. To mimic this process we
assume that the following two conditions should be satisfied
simultaneously,
\begin{eqnarray}
|j|=|B-B^{\prime }| &\geq &j_{\max },  \nonumber \\ B\cdot
B^{\prime } &<&0,  \label{BB'}
\end{eqnarray}
where $B$ and $B^{\prime }$ denote the magnitudes of the magnetic
field in neighboring cells. The conditions like that are supposed
to hold for conventional plane reconnection configuration with an
X-point (Petchek, \cite{Petch}; Syrovatsky, \cite{Syrov}).
Equation~(\ref{BB'}) results in the existence of currents that can
significantly exceed the critical value (more details can be found
in Appendix B).

This is taken into account by requiring that magnetic fields in
adjacent cells have opposite directions, which supposes the
existence of a magnetic null point in between, and corresponds to
an unstable equilibrium favorable to reconnection.
\end{itemize}

This seems to be enough to distinguish between the two dissipation
mechanisms in the framework of the cellular automata model
imposing that the reconnection requires a special configuration
(an X-point for instance), whereas the only criteria for anomalous
resistivity is a current greater than a given threshold. Indeed,
magnetic reconnections corresponds to an important change in
magnetic field's topology, whereas anomalous resistivity does not.
In real physical conditions of the corona both processes can be
present. The threshold current for the anomalous resistivity is in
general supposed to be much larger than for the reconnection under
similar plasma parameters. However our task in this work is
confined in the study of the statistical properties of the energy
dissipation dependence upon parameters. Closer comparison with
real parameters in the corona will be the object of the more
detailed study in the future and comes beyond the scope of this
paper.

The rules that we use to describe the dissipation process are
based on the magnetic field dissipation, that means for us
transformation to the particles energy or heating. For the sake of
simplicity and pure formulation of the problem we consider here
that all the magnetic field energy is transformed completely to
the heating. The rules are as follows: when the current is
annihilated, magnetic field values in both neighboring cells, $B$
and $B^{\prime }$, are replaced by $1/2(B+B^{\prime })$, thus, the
density of magnetic energy dissipated in a single event is given
by $$ \Delta E=\frac{1}{2}\,(B-B^{\prime
})^{2}=\frac{1}{2}\,j^{2}\gtrsim \frac{1}{2}\,j_{\max }^{2}. $$
The procedure modeling the dissipation of currents is the same for
both anomalous resistivity and reconnection. For each time step,
the currents satisfying the dissipation criterion are dissipated
till all the currents become subcritical (or have the same sign in
the case of reconnection). Then, we proceed to the next time step
and switch on the source. Indeed, dissipative processes are
supposed to be faster than the driving ones. The total dissipated
energy is calculated as a sum over all the dissipated currents for
the time step considered.

To compare the effect of a single act of the magnetic field driver
with that of the energy dissipation, one can analyze the spatial
Fourier transform of the energy dissipation and the energy influx,
\begin{eqnarray*}
& I_\mathrm{source}(k) =
\displaystyle
\frac{16}{k_x^2k_y^2}\,
\sin^2\left(\frac{k_x}{2}\right) \sin^2\left(\frac{k_y}{2}\right), \\
& I_\mathrm{diss}(k)  =
\displaystyle
\frac{16j_{\max}^2}{k_x^2k_y^2}\,
\sin^4\left(\frac{k_x}{2}\right)\,\sin^2\left(\frac{k_y}{2}\right),
\end{eqnarray*}
where $k_x,\,k_y\in[-\pi,\pi]$. The last equation holds for the
dissipation of the $y$-component of current. For $x$-component
$k_x$ and $k_y$ should be interchanged.

It is seen that the dependence of these two spectra on $k_x$ are
quite different. For each fixed $k_y$, the energy dissipation
vanishes at $k_x=0$ and increases monotonously with the growth of
$|k_x|$, while the source of the magnetic field is maximum at
$k_x=0$ and decreases with the increase in $|k_x|$. It is worth
noting that for large thresholds one act of dissipation takes
place after a large number of actions of the magnetic field influx
having on each step different phases, which are randomly
distributed. This difference in the action of the magnetic field
source versus dissipation introduces some intermediate scale in
the $k$-space where the action of the source is approximately
compensated by the action of energy dissipation. This
characteristic value of $k$ is closely related to the
characteristic correlation length of the magnetic field spatial
distribution. The growth rate of the magnetic field energy and the
effective damping rate are non-vanishing almost everywhere in the
$k$-space, but the growth rate dominates for small $k$'s (large
spatial scales) while dissipation dominates for larger $k$'s
(smaller spatial scales). Such a situation corresponds in terms of
energy cascade to the ``normal'' direction of the energy flux,
i.e., from large scales to smaller ones. It is also worth
mentioning that using uniform driving results in an important
difference between our model and conventional SOC models, where
the extreme tenuousity of the driving is essential (Sornette et
al., \cite{Sor}). This tenuousity makes the driver nonlocal, in
the sense that it depends on the state of the whole system, as
discussed by Vespigniani \& Zapperi, \cite{ves}.

\subsection{Characteristic spatial scales}  

Dissipation mechanisms and their thresholds depend upon the
parameters of the plasma of solar corona such as background
magnetic field, density, etc. Since our model is aimed at
describing local regions in the corona rather than the corona as a
whole, it is quite natural to assume that the same dissipation
criterion can be applied for each cell of the grid. Thus the
question about the characteristic sizes of the dissipated currents
arises. The observations and theoretical studies show that the
scale of current sheets can be smaller than 1~km. The smallest
scales, of about 10~m, which are considered by Einaudi \& Velli
(\cite{einaudi99}), are associated with the current regions for
the Petchek-type reconnection events. Let us estimate the
characteristic scales of the dissipation events due to anomalous
resistivity. Assuming that the resistivity is provided by the ion
sound instability that have a quite small threshold, we can easily
obtain $$ |\nabla\times\vec{B}|\simeq B/L >
\frac{4\pi}{c}\,n_eec_s, $$ where $L$ is a characteristic width of
the current sheet layer, $c$ is the speed of light, $B$ is the
characteristic magnitude of the magnetic field, $n_e$ is the
plasma density, and $c_s=(T_e/m_i)^{1/2}=(m_e/m_i)^{1/2}\,v_{Te}$
is the ion-sound velocity, $T_e$ is the electron temperature,
$v_{Te}$ is the electron thermal velocity, $m_{e,i}$ are the
electron and ion masses. Then $$ L < \frac{Bc}{4\pi nev_{Te}}
\left(\frac{m_i}{m_e}\right)^{1/2}=
\beta^{-1/2}\,\frac{c}{\omega_{pi}}, $$ where $\omega_{pi}$ is ion
plasma frequency and $\beta$ is the ratio of the kinetic to
magnetic pressure. In the low corona, where $\beta$ is supposed to
be of the order of 1, we have $L\simeq300$~m. This scale is
significantly smaller than the spatial resolution of modern
experimental devices. Moreover, using the angular scattering
measurements of the electron density fluctuations, the smallest
scale that can be resolved in the slow solar wind at $8R_{\odot}$
is 6~km (Woo \& Habbal, \cite{Woo}). Assuming that the linear
structures expand radially as $r$, the structures at $1R_{\odot}$,
where dissipation is supposed to occur, are of the order of 1~km.

Thus, until now, only macroscopic characteristics can be observed.
The statistical microscopic models are aimed at reproducing the
main features of these observations. In this paper we study the
influence of the statistical properties of the magnetic field
source, of the type of dissipation mechanisms and their thresholds
on the macroscopic properties of the total flux of dissipated
energy. This problem is related to the other ones. In particular,
can the local mechanisms of dissipation result in long-range
spatial correlations? In self-organized systems, the appearance of
such correlations give rise to power law distributions of
dissipated energy. Such properties may be caused by specific
features of the source such as the deterministic chaos or
non-Brownian diffusion.

\section{Results}   

Preliminary results concerned with the influence of the type of
magnetic field energy dissipation on statistical properties of the
total radiation energy flux were presented by Podladchikova et al.
(\cite{PKL}). It was shown that the dissipated energy has
approximately normal distribution when the source of the magnetic
field is random and the dissipation is provided by anomalous
resistivity. This result was obtained under the condition that the
current density threshold for the dissipation to occur was
moderate ($j_{\max}=5$). Under the very same conditions, but for
the dissipation due to reconnection, the non-Gaussian energy tails
were observed as well as some other interesting features, in
particular, large-scale spatial correlations of the magnetic
field. This result seems to be quite natural because in this case
some currents may exceed the critical value but not dissipate when
the magnetic fields have the same direction in the neighboring
cells (see Eq.~(\ref{BB'})). The currents may grow up to larger
magnitudes and provide more intensive energy releases in a single
dissipative event and sometimes longer chains of these events. The
large energy events are those that result in significant
deviations from the Gaussian distribution.

\subsection{Transient and stationary states}  

\begin{figure}[!htb]  
\centerline{\epsfig{file=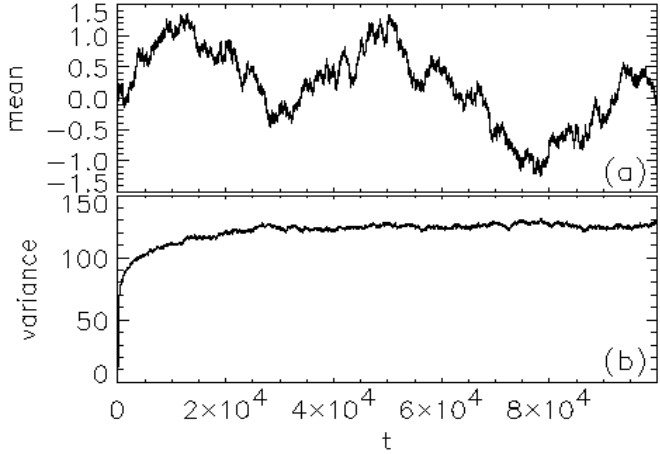,height=6.0 cm,width=8.7 cm}}
\caption{Evolution of mean and variance of the magnetic field.
         For variance, observed are three stages---a linear increase, a
         nonlinear stage, and a stationary state. The results are
         obtained for unipolar random source and anomalous resistivity
         dissipation with a threshold $j_{\max}=30$.
        }
\label{meanvarmono}
\end{figure}

We have studied the dependence of the statistical properties of
our system upon the dissipation threshold $j_{\max}$ ranging
from $0.01$ to 300. To perform statistical analysis correctly,
the averaging procedures should be carried out for the stationary
state of the system. The transient time depends on the threshold.

If the dissipation is absent, the field in each particular cell
would follow a Brownian motion under the influence of the random
source. As a result, the magnetic field in each particular cell,
as well as the average over the whole box, would have a Gaussian
distribution with a growing variance, $\langle B^2\rangle\propto
t$. The currents exhibit the similar behavior, the current
variance growing as $$ \langle j^2 \rangle
(t)\approx2\,\langle\delta B^2\rangle\,t, $$ where for the random
source $\langle\delta B^2\rangle=2/3$. On the average, the current
dissipates for the first time at a moment $$
t_S\simeq\frac{j_{\max}^2}{2<\delta B^2>}. $$ This relationship
also gives a characteristic time between two dissipation events
that occur in the same cell in a stationary state. To obtain a
reliable statistics, the simulation time should significantly
exceed $t_S$. Depending on $j_{\max}$, the simulations performed
have $10^5$--$10^6$ time steps.

Dissipation finally saturates the growth of the variance (see
Fig.~\ref{meanvarmono})b, and a stationary state is reached. The
average over the whole grid magnetic field $B$ undergoes strong
fluctuations but its time average is zero
Fig.~\ref{meanvarmono})a. The average number $n$ of dissipated
currents at each time step can be estimated from the energy
balance considerations. For a single time step, the energy input
on a $N\times N$ grid is $$ \delta E_\mathrm{in} \simeq
N^2\langle\delta B^2\rangle, $$ while the dissipated energy is $$
\delta E_\mathrm{diss} \simeq-n\,\frac{j_{\max}^2}{4}. $$ In the
equilibrium state $\delta E_\mathrm{in}\simeq\delta
E_\mathrm{diss}$, hence
\begin{equation}
\frac{n}{N^{2}}
\simeq\frac{4\langle\delta B^2\rangle}{j_{\max}^2}
\simeq\frac{2}{t_S}.
\label{nN}
\end{equation}
Assuming that these currents are uniformly distributed over the
grid, the characteristic distance between them is $$
l\simeq\frac{N}{\sqrt{n}} = \frac{1}{2}\frac{j_{\max}}
{\sqrt{\langle\delta B^2\rangle}}. $$

\subsection{Grid size effects}   

\begin{figure}[!htb]   
\centerline{\epsfig{file=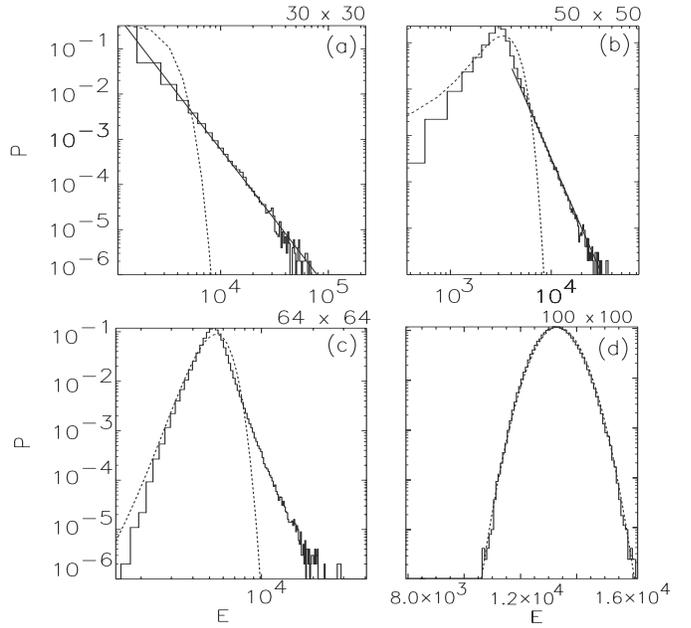,width=8.8cm}}
\caption{Probability distribution function of the dissipated
         energy versus energy released (solid lines). Dotted lines
         represent the best fits by Gaussian distributions. The
         results are obtained for unipolar random source and anomalous
         resistivity dissipation with a threshold $j_{\max} = 5$.
         The grid sizes are the following:
         \textbf{a)} $30\times30$; \textbf{b)} $50\times50$;
         \textbf{c)} $64\times64$; \textbf{d)} $100\times100$.
        }
\label{Egrid}
\end{figure}

To investigate the grid size effects, we take grids consisting of
$30\times30$--$400\times400$ cells, the conditions of anomalous
resistivity dissipation, and $j_{\max}=5$. In this case we obtain the
distributions that are very close to Gaussian for the largest grid.
All the simulations were performed during $10^6$ time steps. The
results are shown in Fig.~\ref{Egrid} for different sizes of the grid.
For the grid $30\times30$, the distribution is nicely fit by a power
law of index $-3.1$ for almost 3 energy decades. As the grid size is
increased, the bulk of the distribution becomes closer to a Gaussian
one, while the high-energy tail retains a power-law shape. For the grid
$50\times50$ the power law shape of the tail is observed only for one
energy decade, the estimate of the index being $-2.9$. For the grid
$64\times64$ the power law tail is again shorter and the index is
approximately equal to $-4.3$. For the grid $100\times100$ the
distribution is practically indistinguishable from Gaussian in the
whole energy range.

Eventually, for large grid size the whole distribution becomes
Gaussian. This means that the character of spatial correlations is
changed. For small grids, spatial correlations decay rather slowly
so that they extend over the whole grid. With the growth of the
grid size, the exponential tails of the spatial correlations
appear yielding the correlation length smaller than the grid size.
The fact that the power-law distributions transform into Gaussian
with the increase in the size of the system considered has already
been observed in some real sandpile experiments (Held et al.,
\cite{Glen}) and in some forest-fire models (Grassberger,
\cite{gra}).

Now we proceed to the discussion of simulation results for various
unipolar and dipolar sources and for both dissipation criteria using
sufficiently large grids to avoid unphysical effects.

\subsection{Random unipolar sources} 

\begin{figure}[!htb]   
\centerline{\epsfig{file=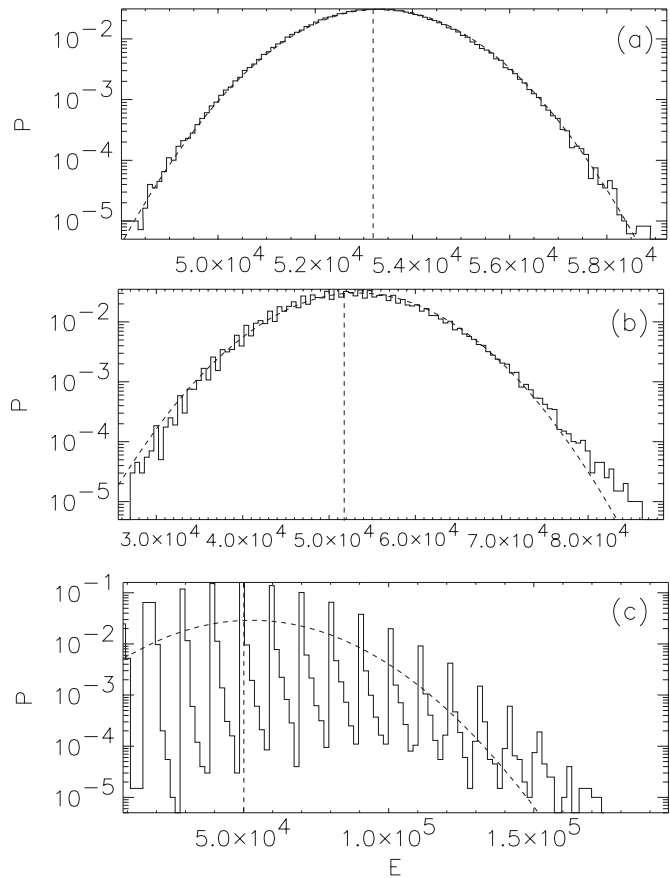,width=8.8cm}}
\caption{Probability distribution function of the dissipated
         energy
         versus energy released (solid lines). Dotted lines represent
         the best fits by Gaussian distributions. The results are
         obtained for unipolar random source and anomalous resistivity
         dissipation. The threshold currents $j_{\max}$ are the
         following:
         \textbf{a)} $j_{\max}=5$; \textbf{b)} $j_{\max}=30$;
         \textbf{c)} $j_{\max}=100$.
        }
\label{pdfa}
\end{figure}

\begin{figure}[!htb]  
\centerline{\epsfig{file=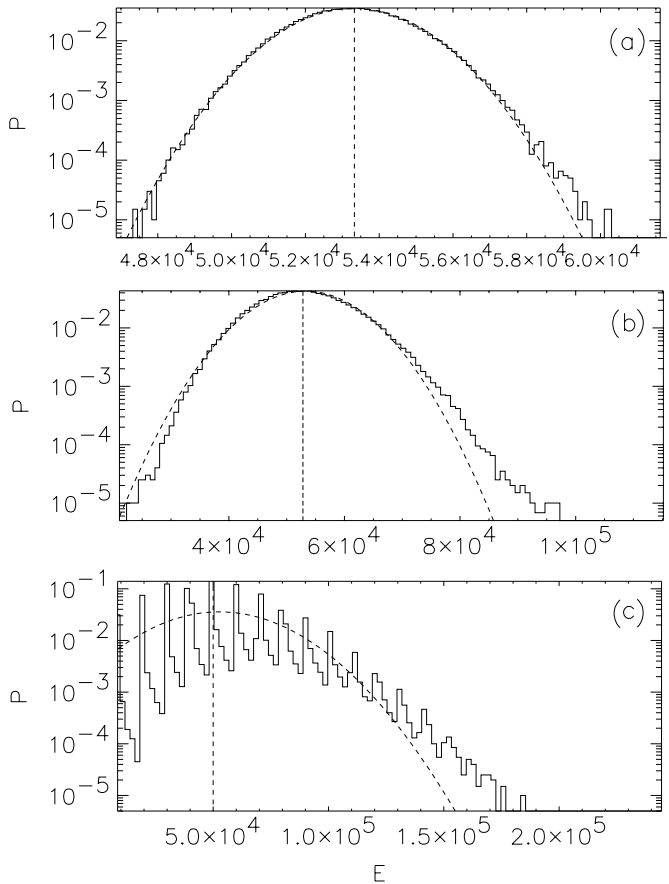,width=8.8cm}}
            \caption{Same as for Fig.~\ref{pdfa}, but for the
            reconnection-like dissipation.}
\label{pdfx}
\end{figure}

Simulations on a $200\times200$ grid were performed for
$2\times10^5$ time steps and both types of dissipation (anomalous
resistivity and reconnection). The distributions of dissipated
energy due to anomalous resistivity with the thresholds
$j_{\max}=5$, 30, 100 are shown in Fig.~\ref{pdfa}.
Figure~\ref{pdfx} represents the same PDF's but for dissipation
provided by reconnection.

The reason for PDF shown in Fig.~\ref{pdfa}c to be multi-extremum
is easy to understand. Since the dissipated energy in each event
is of the form $(j_{\max}+\varepsilon)^2$, where
$0<\varepsilon<2$, for large $j_{\max}$, the peaks separated by
$j_{\max}^2$ appear. Assuming that $\varepsilon$ is a uniformly
distributed random variable, we can get that the width of the peak
number $K$ is of the order of $K\langle \varepsilon^2\rangle$.
Hence, the discrete character of the PDF appears for the smallest
values of the dissipated energy and large $j_{\max}$ (see
Fig.~\ref{pdfa}b--c).

From Figs.~\ref{pdfa} and \ref{pdfx}, one can see that the smaller
current threshold (as compared to the source amplitude) results in
a PDF of dissipated energy that is close to normal. With the
growth of $j_{\max}$, a high energy nonthermal tail appears.
Although the average value of the dissipated energy does not
depend upon $j_{\max}$, the deviations from Gaussian distribution
in the tail increase with $j_{\max}$ and are more pronounced for
dissipation due to reconnection. For the dissipation due to
reconnection, all three distributions have non-Gaussian tails that
can be approximated by power-law. With the increase of $j_{\max}$
the tails become more extended and the index of the power-law
distribution decreases. This signifies that the stronger
deviations from Gaussian distribution appear for larger values of
the threshold.

The formation of such a tail is not related to the increase of the
correlation length, because the largest magnetic field correlation
length is about 20 for moderate threshold, $j_{\max}=5$, in the
case of reconnection-like dissipation. Let us also notice that the
average correlation length is smaller when anomalous resistivity
dissipation is considered and decreases as $j_{\max}$ increases
(Podladchikova et al., \cite{PKL}).

The average number of dissipated currents decreases with the
increase of the threshold. Indeed, it easily seen from
Eq.~(\ref{nN}) that for $j_{\max}=1$ we have $$ \frac{n}{N^2} =
\frac{4}{3}, $$ i.e., each current is dissipated at each time step
and dissipative events occur almost everywhere on the grid thereby
creating long range correlations. For large $j_{\max}$, the ratio
$n/N^2$ becomes small, e.g., for $j_{\max}=100$ we have
$n/N^2\simeq3\cdot10^{-4}$ that corresponds to the uncorrelated
dissipative events with relatively small overlapping.

Thus, for large $j_{\max}$ the dissipation is much faster than the
action of the magnetic field source, i.e., the conditions of time scale
separation are better satisfied.

\begin{figure}[!htb]  
\centerline{\epsfig{file=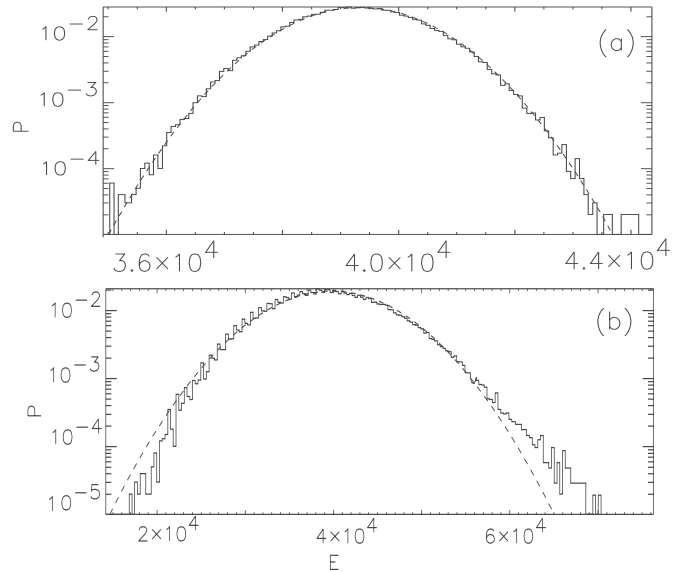,width=8.8cm}}
\caption{Probability distribution function of the dissipated
energy versus energy released (solid lines). Dotted lines
represent the best fits by Gaussian distributions. The results are
obtained for Ulam map source and $j_{\max} = 30$.  . The
dissipation laws are the
         following:
         \textbf{a)} anomalous resistivity dissipation;
         \textbf{b)} reconnection-type disispation.
        }
\label{puax}
\end{figure}

\subsection{Ulam map source}  

Despite the deterministic nature of the magnetic field increment
$\delta B$, without dissipation the magnetic field in each cell
exhibits a Brownian motion with Gaussian statistics after a large
number of steps. Indeed, each $\delta B$ has the same PDF with a
finite variance. Thus, the dependence of the PDF upon the critical
current and the dissipation mechanism have manifested itself in
similar tendencies as random sources do. The PDF's for moderate
critical current, $j_{\max}=5$, are close to Gaussian, for
$j_{\max}=30$ the small deviations from Gaussian distribution
begin to appear for high energies, and for $j_{\max}=100$ there is
clear evidence of the presence of the high energy suprathermal
tail. The tail is much better pronounced for the dissipation due
to reconnection (Figure~\ref{puax}).

\begin{figure}[!htb]  
\centerline{\epsfig{file=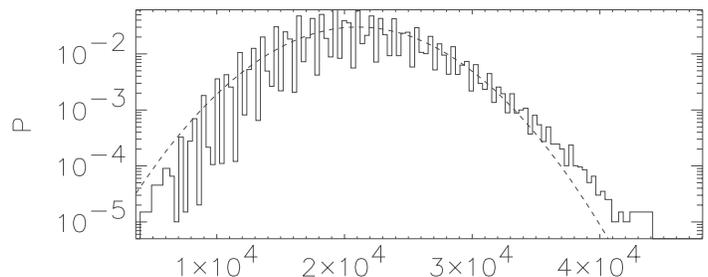,width=9.25cm}}
\caption{Probability distribution function of the dissipated
energy
         versus energy released (solid lines). Dotted lines represent
         the best fits by Gaussian distributions. The results are
         obtained for Geisel map source and anomalous resistivity
         dissipation. Threshold current $j_{\max}= 30$.}
\label{pdfga}
\end{figure}

\begin{figure}[!htb]  
\centerline{\epsfig{file=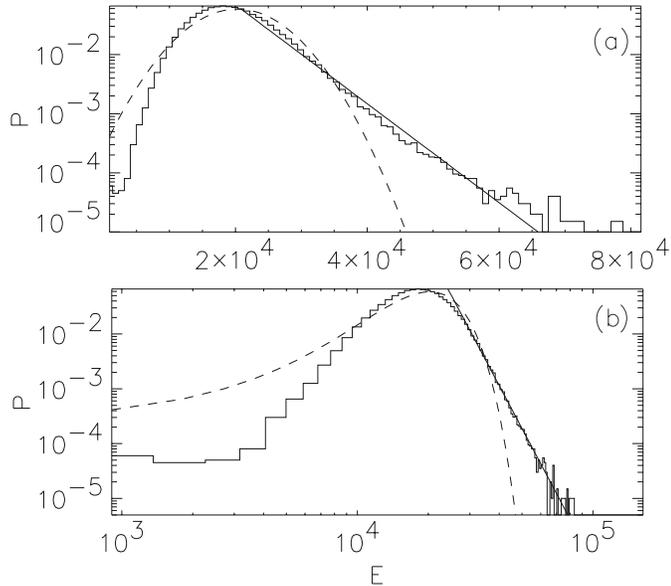,width=8.8cm}}
\caption{Probability distribution function of the dissipated energy
         versus energy released (solid lines). Dotted lines represent
         the best fits by Gaussian distributions. The results are
         obtained for Geisel map source and dissipation due to
         reconnection. Threshold current is $j_{\max}=30$.}
\label{pdfgx}
\end{figure}

\subsection{Geisel map lattice}  

Another source of the magnetic field is provided by the Geisel
map. It was used as a source of the field rather than the
increment of the field. As it was discussed above, this map
exhibits the time evolution in each cell quite similar to
anomalous (non-Brownian) diffusion. The growth of the magnetic
field variance is slower than for Brownian motion. Such a behavior
of the walker would be called subdiffusive. As a result, the
variance is smaller than for the random source. This causes some
differences observed in the behavior of the PDF of the dissipated
energy. If the dissipation is provided by anomalous resistivity,
the PDF's obtained with this map and the random source are quite
similar except the width of the distribution. Figure~\ref{pdfga}
represents the PDF obtained for Geisel map and $j_{\max}=30$.

The smaller current threshold (as compared to the source
amplitude) results in a PDF of dissipated energy that is close to
normal. With the growth of $j_{\max}$, a high energy nonthermal
tail appears. The deviations from Gaussian distribution in the
tail increase with $j_{\max}$. With the increase of $j_{\max}$ the
tails become more extended.

In this case the range of energies is a slightly less than for the
Ulam map, the distribution being quite jagged, all the other
features of the distributions seem to be similar.

Figure~\ref{pdfgx} shows the PDF obtained for the same Geisel
map source, dissipation due to reconnection, and $j_{\max}=30$. A
pronounced tail of the distribution is evidenced thereby
confirming the tendencies already observed for other types of
sources. A difficulty that one encounters when analyzing such
PDF's is illustrated in Fig.~\ref{pdfgx}. In Fig.~\ref{pdfgx}a,
the PDF is represented in semi-logarithmic scale and the
distribution seems to have an exponential tail. In
Fig.~\ref{pdfgx}b the same distribution is shown in the log-log
representation, where the same tail can be treated as a
power-law-type distribution with the index $-3.1$. This problem is
related to the fact that the tail extends over a rather short
range of about one energy decade. This problem will be treated in
more detail elsewhere (Podladchikova et al. \cite{korova}).

\begin{figure}[!htb]  
\centerline{\epsfig{file=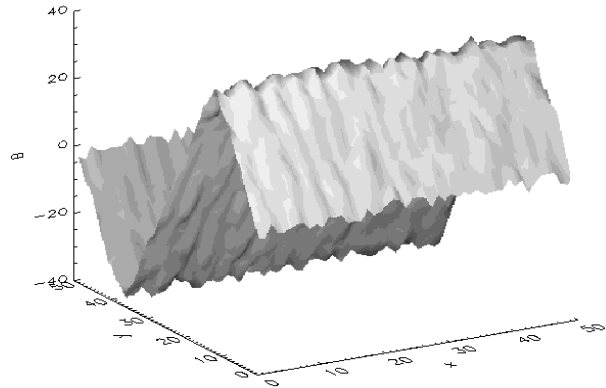,width=8.cm, height=5.3cm}}
\caption{Magnetic field structure observed in the simulations with
the dipolar random source, anomalous resistivity dissipation, and
a threshold current $j_{\max}=5$.} \label{bipff}
\end{figure}

\begin{figure}[!htb] 
\centerline{\epsfig{file=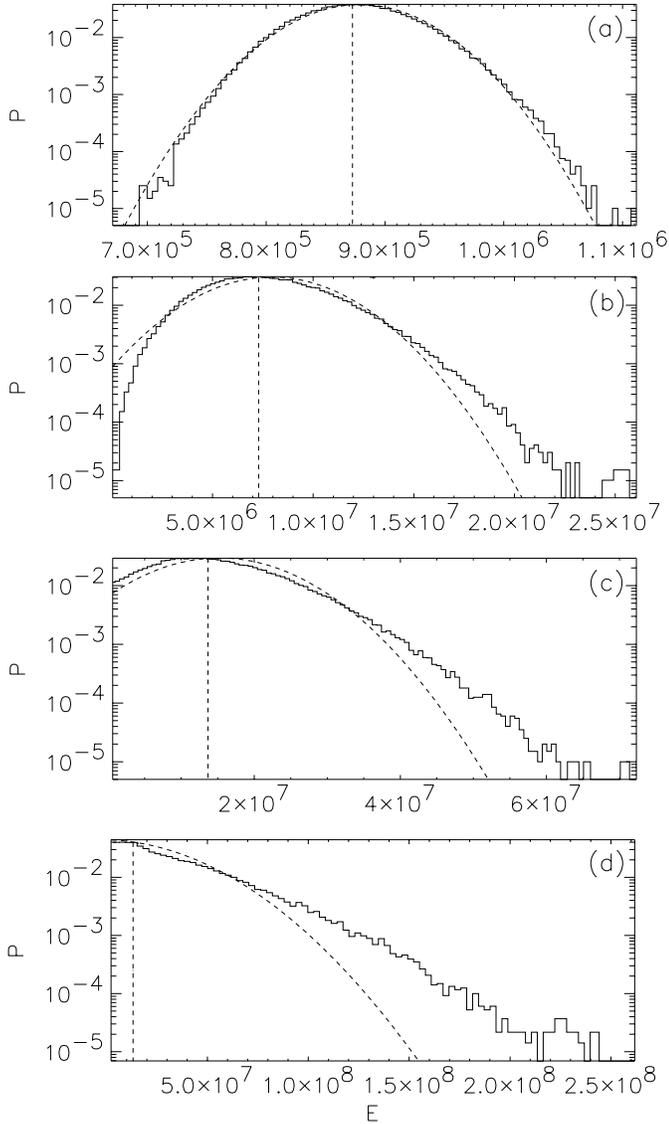,width=8.8cm}}
\caption{Probability distribution function of the dissipated
energy versus the energy released (solid lines). Dotted lines
represent the best fits by Gaussian distributions. The results are
obtained for dipolar random source and anomalous resistivity
dissipation. The duration of calculations was $2\times10^5$ time
steps. The threshold currents $j_{\max}$ are the following:
\textbf{a)} $j_{\max}=5$; \textbf{b)} $j_{\max}=50$; \textbf{c)}
$j_{\max}=100$; \textbf{d)} $j_{\max}=230$.} \label{Apdfb}
\end{figure}

\begin{figure}[tbh]   
\centerline{\epsfig{file=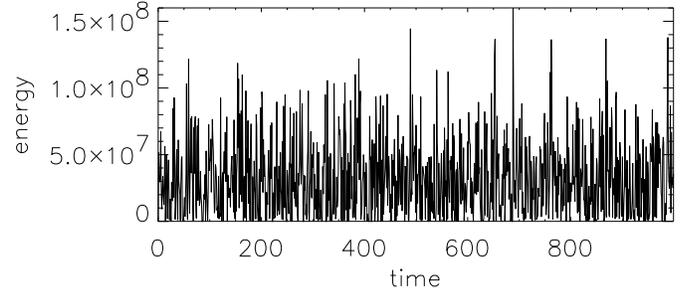,width=8.8cm}} \caption{Fragment of
time series of the dissipated energy, for $j_{\max}=230$, dipolar
random source, anomalous resistivity dissipation.} \label{E2301}
\end{figure}

 \subsection{Dipolar source}  

When the effects due to the bi-polarity are modeled, the grid is
split into two equal parts. For one half of the grid, the value of
the magnetic field increment is chosen randomly from the set
$\{-1.5, -0.5, 0.5\}$, for another half the set $\{1.5, 0.5,
-0.5\}$ is used. The global inhomogeneity of the dipolar source
results in the formation of a strongly localized current layer in
the transition region between the two parts of the system. The
grid size is $100\times100$.

The dissipation provided by reconnection is not always able to
stabilize the field growth. However, for dissipation due to
anomalous resistivity a stationary state always exists. Further we
consider only the case with anomalous resistivity dissipation.

The energy of the dipolar structure grows faster than in the
unipolar case, and a stationary state is quickly reached. The
dipolar structure of the magnetic field is presented in
Fig.~\ref{bipff}. This structure is stationary. Therefore the
spatial correlations of magnetic field are long and
quasi-stationary.

Such a system differs from the unipolar one with the random source. We
observe, as could be expected, that the dissipation events are more
intense within the transition region where the magnetic field changes
its sign. In addition, in the dipolar case the position of the PDF
peak value on the energy axis is significantly shifted towards high
energies with the increase of the threshold. Such a property can be
explained by the presence of the localized current layer in the
vicinity of the neutral sheet. The characteristic dependence of the
PDF's upon the dissipated energy in dipolar case displays a
supra-thermal high energy tails which were not observed in the unipolar
case with anomalous resistivity (see Fig.~\ref{Apdfb}). The
characteristic indices for $j_{\max}=50$, 100, and 230 are
approximately equal to $-1.7$, $-1.9$, and $-2.3$, respectively.  For
$j_{\max}=230$, the time series of the dissipated energy is shown in
Fig.~\ref{E2301}.

\section{Discussion and conclusions}  

To study coronal heating due to dissipation of small-scale current
layers, we have performed a statistical analysis of a simple
cellular automata model. Its principal difference with previous
ones is that the system is driven by small scale homogeneously
distributed sources acting on the entire grid for each time step.
The idea to consider small scale sources is similar to the one
proposed by Benz \& Krucker that the heating occurs on the level
of the chromosphere, thus the magnetic field structures,
dissipation of which supplies the energy for the heating, are also
of a small scale.

The magnetic field sources we use are either of the following
types:

\begin{itemize}
\item  random source,

\item  deterministic chaotic map (Ulam map), both for the magnetic field increment.

\item  Geisel map (coupled map lattice) that describes the time evolution of
the magnetic field in each cell similar to anomalous
(non-Brownian) diffusion.
\end{itemize}

We consider two mechanisms of small scale dissipation to occur:

\begin{itemize}
\item the first is used to model anomalous resistivity dissipation,

\item the second is used to model the local reconnection.
\end{itemize}

The first one is supposed to be similar to Joule dissipation but
in collisionless plasma. It relies on the criterion that any local
current whose magnitude exceeds a pre-determined threshold value
must dissipate.

In the second case we assume that the dissipation occurs when the
above mentioned condition is satisfied (this does not mean that
the threshold in real physical system should be the same in two
cases) and, in addition, the magnetic fields vectors in
neighboring cells should have opposite signs. This mimics the
presence of the null point of the magnetic field in the
reconnection-type configurations.

The characteristic under investigation is the total energy of all
simultaneously dissipated currents.

Our observations can be summarized as follows.

Small scale magnetic field sources and localized energy
dissipation mechanisms can result in large scale correlations of
the magnetic field. In our calculations, the energy sources
dominate in larger scales while the dissipation in the smaller
ones. Thus, the system behaves as having the energy cascade from
large scales to smaller ones. However, during the evolution of the
system, large scale correlations are formed with a characteristic
length significantly larger than that of the source and
dissipation.

For all three types of the magnetic field sources considered, the
obtained PDF of the dissipated energy is close to Gaussian
distribution when relatively small threshold currents,
$j_{max}<10$, are chosen. The dependence of the PDF of the
dissipated energy upon statistical properties of the source for
all three types of the sources considered is rather weak and
requires an additional thorough analysis to be performed.

In the case of reconnection type dissipation, the deviations from
Gaussian distribution are stronger than for anomalous resistivity
dissipation. For large values of the threshold current density, we
observe a high-energy suprathermal tail which have the shape
similar to power-law distribution. The time-averaged spatial
correlations are exponentially decaying. In this sense our model
can not be considered as a self-organized critical system.
However, sometimes long range (power-law type) correlations are
observed.

To model a dipolar global magnetic field structure we used
inhomogeneous distribution of the magnetic field source. It has
small average positive value in the one part of the grid and
negative in another with the summary average equal to zero. The
system does not reach any stationary state in the case of the
reconnection type dissipation. In the case of anomalous resitivity
it saturates and the stationary state is established. In this last
case, the deviations from the Gaussianity become stronger with the
increase of the threshold. This effect is associated with the
decrease of the thickness of the current layer where the
dissipation is concentrated.

The statistical analysis of frequency distribution of such heating
events shows always larger absolute value of the exponent for the
more energetic reconnection produced events than for the anomalous
resistivity heating events. This tendency seems to be similar to
the one found by Benz \& Krucker, \cite{Benz98} who studied the
emission measure increases. They pointed out that the statistical
properties of the faint events that occur in the intra-cell
regions of the quiet Sun manifest quite small deviations from
Gaussian distributions, while supposed nanoflares that are
associated with the network boundaries have more prominent
enhancements and stronger deviations. Taking this into account we
come to the conclusion that the quantitative difference between
faint and strong heating events reported in (Krucker \& Benz,
\cite{Krucker00}) can probably be explained by two different
mechanisms of magnetic field dissipation that we used in our
model.

However, our current work represents only the first step in the
development of a model. In particular, many elements of its should
be better adapted to real physical processes. Further effects
should be included for direct comparison with the observations,
such as more detailed description of respective time scales of the
instabilities or particle acceleration, thermal/non-thermal
processes or outgoing flows escaping from the reconnection region.
For instance one can assume that in the case of the anomalous
resistivity the dissipation of the current can be partial in spite
of the complete, the two dissipation mechanisms can be combined,
anomalous resistivity can co-exist with the reconnection. One can
more carefully take into account the energy re-distribution
between different particle species as we have noted in the
Appendices. Thus, on this stage we can only conclude that some
tendencies of our model seem to be in a good qualitative agreement
with the experimental ones.

\begin{acknowledgements} The authors would like to acknowledge
Professor A. Benz for fruitful comments that helped to improve
this paper. The authors are thankful to V.~Lobzin, T.~Dudok de
Wit, S. Koutchmy and S.M.~Levitsky for fruitful and useful
discussions. B.~Lefebvre is grateful to JSPS for financial
support. O.~Podladchikova is grateful to French Embassy in Ukraine
for the financial support.
\end{acknowledgements}

\appendix
\section{Anomalous resistivity}

In this appendix, we explain in more details the physics of the
''anomalous resistivity'' mentioned in our paper. The idea is
based on the possibility of the exchange of the pulse and energy
between the electrons and the ions due to the instability
development and the appearance of the turbulent state in the
plasma. The ''conventional'' expression for the conductivity reads
$$ \sigma =\left( ne^{2}\right) /\left( m_{e}\nu \right) , $$
where $n$ is the plasma density, $e,m_{e}$ - electron charge and
mass respectively, $\nu $ is the collision frequency of the
electrons with ions. In collisionless plasma the collision
frequency is negligible and the collisional conductivity is
infinite. However the electrons (who carry the major part of the
electric current) can excite collective oscillations of the
electrons as well as ions, and transfer part of their impulse and
energy to these oscillations. This results to an anomalous loss of
the electron pulse and energy, and, consequently, to the decrease
of their directed velocity, i.e. the decrease of the current. Such
a process can be characterized by an ''effective collision
frequency'' $\nu _{eff}$. To calculate this characteristic
frequency one can consider the pulse conservation law in the
system that consists of electrons and waves. To this end one
should evaluate this effect as the action of the friction force
that decelerates the electron flow. The pulse loss per unit time
can be written as follows: $$ \nu
_{eff}m_{e}n\mathbf{U_{d}}=-\mathbf{F_{fr}} $$ where $U_{d}$ is
the directed velocity of the electron flow that carries the
current, and $F_{fr}$ the friction force that acts on the
electrons. The same decrease of the electron pulse can be
estimated as the increase of the pulse of the waves emitted by the
electrons taking into account that the change of the pulse of the
waves is described by the following expression: $$
\frac{dP_{w}}{dt}=2\int \gamma _{k}W_{k}\frac{\mathbf{k}}{\omega _{k}%
}\frac{d^{3}k}{\left( 2\pi \right) ^{3}} $$ where $\gamma _{k}$ is
the linear increment of the instability and $W_{k}$ is the
spectral energy density of waves. The assumption that the pulse of
the system consisting of electrons and waves is conserved results
in the equality of these two expressions, thus the effective
collision frequency can be defined as: $$ \nu
_{eff}=\frac{2}{m_{e}nU_{d}^{2}}\int \gamma _{k}W_{k}\frac{\left(
\mathbf{U_{d}}\mathbf{k}\right) }{\omega _{k}}\frac{d^{3}k}{%
\left( 2\pi \right) ^{3}}. $$ Thus the estimate of the level of
the wave turbulence taking into account the nonlinear saturation
mechanism allows to solve the problem.

When the current flows in the direction perpendicular to the
magnetic field two major types of the instabilities that give rise
to the anomalous resistivity are modified Buneman instability and
the instability with respect to the generation of the so called
electron-acoustic modes. The final effect of the anomalous
turbulent resistivity is the transfer of the energy from the
electrons to ions, because the damping of these waves takes place
mainly due to their interaction with ions.

In the particular case of a current perpendicular to the magnetic
field, two major types of instabilities that result in the
anomalous resistivity are modified Buneman instability and the
instability which generates the so called electron-acoustic waves.

An important physical characteristics of the anomalous resistivity
phenomenon is the ratio of the energy dissipated by ions and to
the energy dissipated by the electrons. If the characteristic
drift velocity of the electrons with respect to ions is
$\mathbf{V_{d}}$, and we have the knowledge of the spectral
characteristics of the wave spectrum excited, let us assume, for
instance that the characteristic frequency of the waves excited
due to the instability is $\omega _{\mathbf{k}}$, and that
characteristic wave vector of the unstable waves is $\mathbf{k}$,
then this ratio can be estimated by: $$
\frac{dE_{e}}{dt}/\frac{dE_{i}}{dt}\sim \frac{\int \gamma _{\mathbf{k%
}}W_{\mathbf{k}}\frac{\left( \mathbf{k}\mathbf{V_{d}}%
\right) }{\omega _{\mathbf{k}}}d^{3}\mathbf{k}}{\int \gamma
_{\mathbf{k}}W_{\mathbf{k}}d^{3}\mathbf{k}}\sim
\frac{\left( \mathbf{k}\mathbf{V_{d}}\right) }{\omega _{%
\mathbf{k}}} $$ for typical instabilities, where this ratio is
approximately 1.

So, the final result of the anomalous resistivity is quite similar
to the Joule heating of the ion component of the plasma. The
energy dissipation that heats the plasma can be represented as: $$
Q=j^{2}/\sigma _{eff} $$ This phenomenon was observed
experimentally in the laboratory plasma (Eselevich et al.,
\cite{eselevich}).

\section{Reconnection}

There are several differences between reconnection process and
anomalous resistivity. One of them is related with the change of
the magnetic field topology in the first case while in the second
there are only quantitative variations of the basic
characteristics of the magnetic field configuration.

Another difference is that while the diffusion processes in the
first case provide heating (comparable for ions and electrons, see
appendix), reconnection converts magnetic energy mostly into
particle (mainly ions) acceleration. Then energetic beams may
provide heat, but this is only an indirect consequence of the
reconnection. One more but quite important for observations is a
difference in time scales. The reconnection is supposed to be a
rapid energy release while the anomalous resistivity is relatively
slow diffusive process.

Moreover, as it was shown analytically and in computer
simulations, the reconnection process can under certain conditions
look like an explosive event. During the reconstruction of the
magnetic field topology the component of the magnetic field
perpendicular to the background magnetic field and to the
direction of the current can grow so rapidly that its variation in
time is quite similar to an explosion. The spatial temporal
dynamics of the magnetic field can be described by the following
expression: $$ \mathbf{B}=B_{0x}\tanh \left( \frac{z}{L}\right)
\mathbf{e}_{x}+B_{z}\left( t\right) \sin \left( kx\right)
\mathbf{e}_{z}, $$ where $$ B_{z}\left( t\right)
=\frac{B_{0z}}{1-t/\tau _{expl}}. $$ (see Galeev, \cite{galeev2}
for more details). Here $B_{0x}$ is the magnitude of the
background surrounding magnetic field, that is supplied by the
current carried along the $y$ axes, $\tau _{expl}$ is the
characteristic time of the magnetic and electric fields
variations. It has opposite sign from two sides of the current
layer. $B_{z}$ is the initial amplitude of the perturbation of the
normal component of the magnetic field that is supposed to be
periodic along the $x$ axes with the wavelength $\lambda = {2
\pi}\mathbf{k})$. Its time dependence represents the growth of the
amplitude up to infinity for the finite time. However this formal
solution is valid only when the amplitude of this perturbation is
smaller than the background field, thus the singularity is not so
important. This explosive growth of nonlinear perturbations
results in the similar increase of the inductive electric field.
It does not mean that the electric and magnetic fields become
infinite, just the nonlinear stage develops in such a rapid way
till the normal component of the magnetic field becomes comparable
with the $B_{0x}$. This process results in the rapid acceleration
of electrons and ions by the inductive electric field in the
region where the particles are unmagnetized.

Another observational feature of the reconnection process consists
in the presence of the macroscopic fluxes around the reconnection
site with the characteristic velocities of the order of the
Alfv\'en speed estimated using background magnetic field in the
vicinity of it. These flows can give rise to the replenishing of
the lower density regions by the material from the reconnection
site if the density there is larger or even comparable with the
density of the surrounding plasma. A part of the energy released
is supposed to be transformed into the kinetic energy of
accelerated charged particles. It is known that the energy
transferred to particles growth with the increase of their mass as
$M^{1/3}$, the heaviest ones being the most effectively
accelerated (Vekstein \& Priest, \cite{veks}). In their turn,
these particles can generate the electromagnetic radiation
providing experimental signatures of the heating events.

\end{document}